\documentclass[iop, twocolumn]{aastex6}
\hypersetup{citecolor=blue,urlcolor=red}
\usepackage{amsmath}
\usepackage{url}
\usepackage{enumerate}
\usepackage{multirow}

\shorttitle{PSR B1534+12}
\shortauthors{Wang et al.}

\begin{document}
\title{The two emission states of PSR B1534+12}
\email{wangjingbo@xao.ac.cn}
\author
{S. Q. Wang\altaffilmark{1,2,3,4}, G. Hobbs\altaffilmark{3}, J. B. Wang\altaffilmark{1,5,6}, R. Manchester\altaffilmark{3}, N. Wang\altaffilmark{1,5,6}, S. B. Zhang\altaffilmark{7,2,3}, Y. Feng\altaffilmark{8,2,3}, W. -Y. Wang\altaffilmark{9,2}, D. Li\altaffilmark{8,2,10}, S. Dai\altaffilmark{3}, K. J. Lee\altaffilmark{11}, S. J. Dang\altaffilmark{1,2,12}, L. Zhang\altaffilmark{8,2,3}}

\altaffiltext{1}{Xinjiang Astronomical Observatory, Chinese Academy of Sciences, 150 Science 1-Street, Urumqi, Xinjiang 830011, China; wangjingbo@xao.ac.cn}
\altaffiltext{2}{University of Chinese Academy of Sciences, Beijing 100049, China}
\altaffiltext{3}{CSIRO Astronomy and Space Science, PO Box 76, Epping, NSW 1710, Australia}
\altaffiltext{4}{CAS Key Laboratory of FAST, National Astronomical Observatories}
\altaffiltext{5}{Key Laboratory of Radio Astronomy, Chinese Academy of Sciences, 150 Science 1-Street, Urumqi, Xinjiang, 830011, China}
\altaffiltext{6}{Xinjiang Key Laboratory of Radio Astrophysics, 150 Science1-Street, Urumqi, Xinjiang, 830011, China}
\altaffiltext{7}{Purple Mountain Observatory, Chinese Academy of Sciences, Nanjing 210008, China}
\altaffiltext{8}{National Astronomical Observatories, Chinese Academy of Sciences, A20 Datun Road, Chaoyang District, Beijing 100101, China}
\altaffiltext{9}{Key Laboratory for Computational Astrophysics, National Astronomical Observatories, Chinese Academy of Sciences, 20A Datun Road, Beijing 100101, China}
\altaffiltext{10}{NAOC-UKZN Computational Astrophysics Centre, University of KwaZulu-Natal, Durban 4000, South Africa}
\altaffiltext{11}{Kavli Institute for Astronomy and Astrophysics, Peking University, Beijing 100871}
\altaffiltext{12}{School of Physics and Electronic Science, Guizhou Normal University, Guiyang, 550001}

\begin{abstract}

We have observed PSR~B1534+12 (J1537+1155), a pulsar with a neutron star companion, using the Five-hundred-meter Aperture Spherical radio Telescope (FAST). We found that this pulsar shows two distinct emission states: a weak state with a wide pulse profile and a burst state with a narrow pulse profile.   The weak state is always present. We cannot, with our current data, determine whether the pulse energy of the weak state follows a normal or a log-normal distribution. The burst state energy distribution follows a power-law. 
The amplitude of the single pulse emission in the burst state varies significantly; the peak flux intensity of the brightest pulse is 334 times stronger than that of the average pulse.  We also examined the timing precision achievable using only bright pulses, which showed no demonstrable improvement because of pulse jitter and therefore quantified the jitter noise level for this pulsar.
\end{abstract}


\section{INTRODUCTION}

Pulsars are fast rotating and highly magnetised neutron stars. 
They exhibit diverse emission properties.  
For instance, pulsars can suddenly switch off (known as pulse nulling or intermittency; e.g., \citealt{Backer70}) or the profile discretely changes (known as pulse mode changing; e.g., \citealt{Bartel82}). 
The nulling timescale is typically in the range of several pulse periods to hours ~\citep{Wang07}, whereas the intermittency timescale is from many days to years~\citep{Kramer06,Wang20}.
Observed links between the nulling and mode changing phenomena suggest that they are the two manifestations of the same phenomenon~\citep{Wang07}. 
To date, both nulling and mode changing have been detected in more than 200 normal pulsars.

A single pulse for some pulsars can be tens or hundreds of times brighter than the average pulse. This rare phenomenon is known as a giant pulse and such pulses have been detected in sixteen pulsars: eleven normal pulsars and five millisecond pulsars~\citep{Ershov03, Hankins03, Johnston03, Kuzmin04, Soglasnov04, Knight05, Kuzmin06, Tsai15}.  The giant pulses are often associated with high energy emission~\citep{Romani01}.  The pulse energy  for the giant pulses generally follows a power-law distribution in contrast to the pulse energy for normal pulses which follows a normal or log-normal distribution~\citep{Karuppusamy10,Mickaliger2018}. 
A similar phenomenon is the giant micro-pulses, in which the flux density is significantly greater than typical value at specific pulse phases, but the overall integrated flux density of the profile remains approximately constant ~\citep{Kramer02}. This phenomenon has been detected in some young pulsars, such as the Vela pulsar \citep{Kramer02}, PSR~B1706$-$44 \citep{Johnston02} and PSR J0901$-$4624 \citep{Raithel15}.  The energy density of these micro-pulses also follow a power-law distribution.

The single pulse variations influence the timing precision achievable for a given pulsar~\citep{Cordes10,Shannon14} and provide a fundamental limit on the achievable timing precision on a short time-scale.   This is known as ``pulse jitter''.

PSR~B1534+12 is a millisecond pulsar with a 37.9\,ms period in orbit with a neutron star companion with a 10.1\,hr orbital period. This pulsar was discovered using the Arecibo 305\,m radio telescope in the 430\,MHz band~\citep{Wolszczan91}.   Timing analysis of this pulsar has produced precise measurements of five post-Keplerian parameters, making it valuable in testing gravitational theories~\citep{Stairs02, Fonseca14}. \citet{Sallmen95} described the single pulse statistics for this pulsar at 430\,MHz using the Arecibo telescope and detected a few narrow single pulses with a characteristic width of $\sim$160\,$\mu$s (see also \citealt{Sallmen98}). PSR~B1534+12 is the first double neutron star binary pulsar that shows this phenomenon.

In this paper we have continued this earlier work, but with the highly sensitive Five-hundred-meter Aperture Spherical radio telescope \citep[FAST,~][]{nan11} radio telescope. 
In Section 2 of this paper, we describe our observation of PSR~B1534+12. In Section 3, we present our analysis of the single pulses detected during this observation.   We summarise our results in Section 4.

\section{OBSERVATIONS AND DATA PROCESSING}

FAST is equipped with an L-band array of 19 feed horns \citep{li18} that covers 1.05 to 1.45 GHz and was installed in May 2018~\citep{Jiang19}. In this work, the central beam of that receiver was used to observe PSR~B1534+12 on 2019 July 27 for 16 minutes. A total of 26115 pulses were obtained during that observation with a time resolution of 49.152\,$\mu$s and a channel bandwidth of 0.122\,MHz.  The data were recorded in a search mode PSRFITS file~\citep{Hotan04}.

To analyze the data set, we used {\sc dspsr}~\citep{Hotan04} to extract individual pulses according to the timing ephemeris provided by the pulsar catalog (PSRCAT; \citealt{Manchester05}) (note that we use the \textsc{-K} option in {\sc dspsr}, which removes the inter-channel dispersion delays\footnote{FAST does not have the signal processing capability to provide coherently dedispersed search mode data streams.  The dispersive delay for the pulsar across a single frequency channel, of 0.122\,MHz is 6 $\mu$s.}). We averaged the time samples to produce 512 phase bins for each pulse profile. The observation was affected by radio-frequency interference (RFI).  We first automatically flagged the RFI and removed 5 percent of the band-edges using the {\sc psrchive} software package~\citep{Hotan04}. We completed the RFI flagging using the interactive  software package {\sc pazi} that allowed us to identify any remaining frequency channels affected by RFI. In total we removed 1186 frequency channels corresponding to 29\% of the total. Further details of the RFI environment around FAST are provided in Section 4 of~\citet{Jiang19}.

Calibration files, in which a switched calibrator source was used, were also recorded and subsequently folded at the calibration pulse period of 2.01326592\,s.  The pulsar observation was subsequently calibrated using the {\sc psrchive} program {\sc pac} to flatten the bandpass and to transform the polarisation products to Stokes parameters.  Topocentric times of arrival (ToAs) were obtained by cross correlating the mean pulse profile with a noise-free template using {\sc psrchive}. Timing residuals were formed using the {\sc tempo2} software package ~\citep{Hobbs06}.

\section{RESULTS}

\subsection{The two emission states}

\begin{figure*}
\centering
\includegraphics[width=160mm]{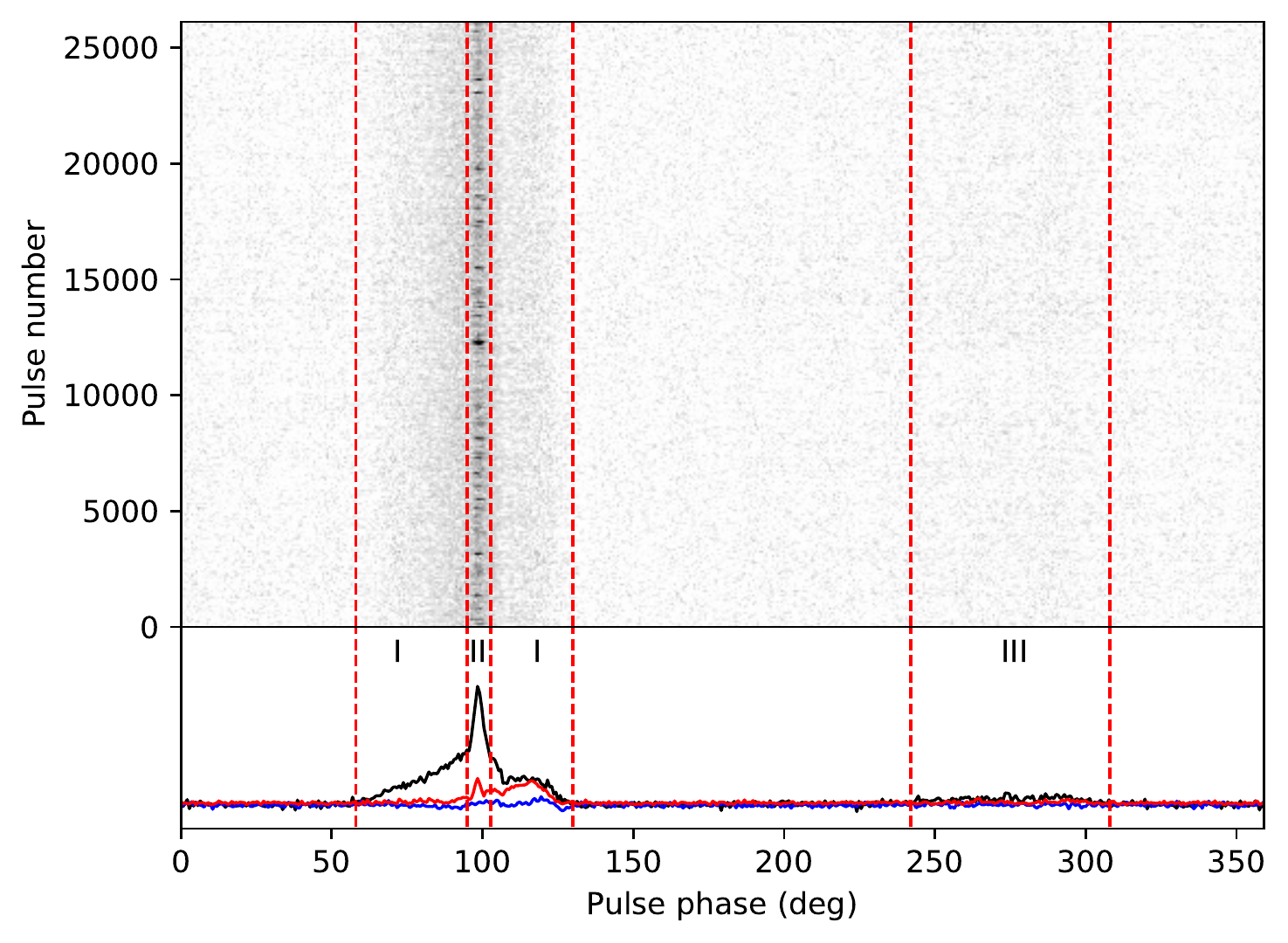}
\caption{Individual pulses sequence and the accumulated pulse profiles for PSR~B1534+12. The average pulse profile is shown in the bottom panel.  The main pulse is divided into two components by the vertical red dashed lines, labelled ``I'' and ``II''. The interpulse is labelled as ``III''.}\label{gray}
\end{figure*}

\begin{figure}
\centering
\includegraphics[width=80mm]{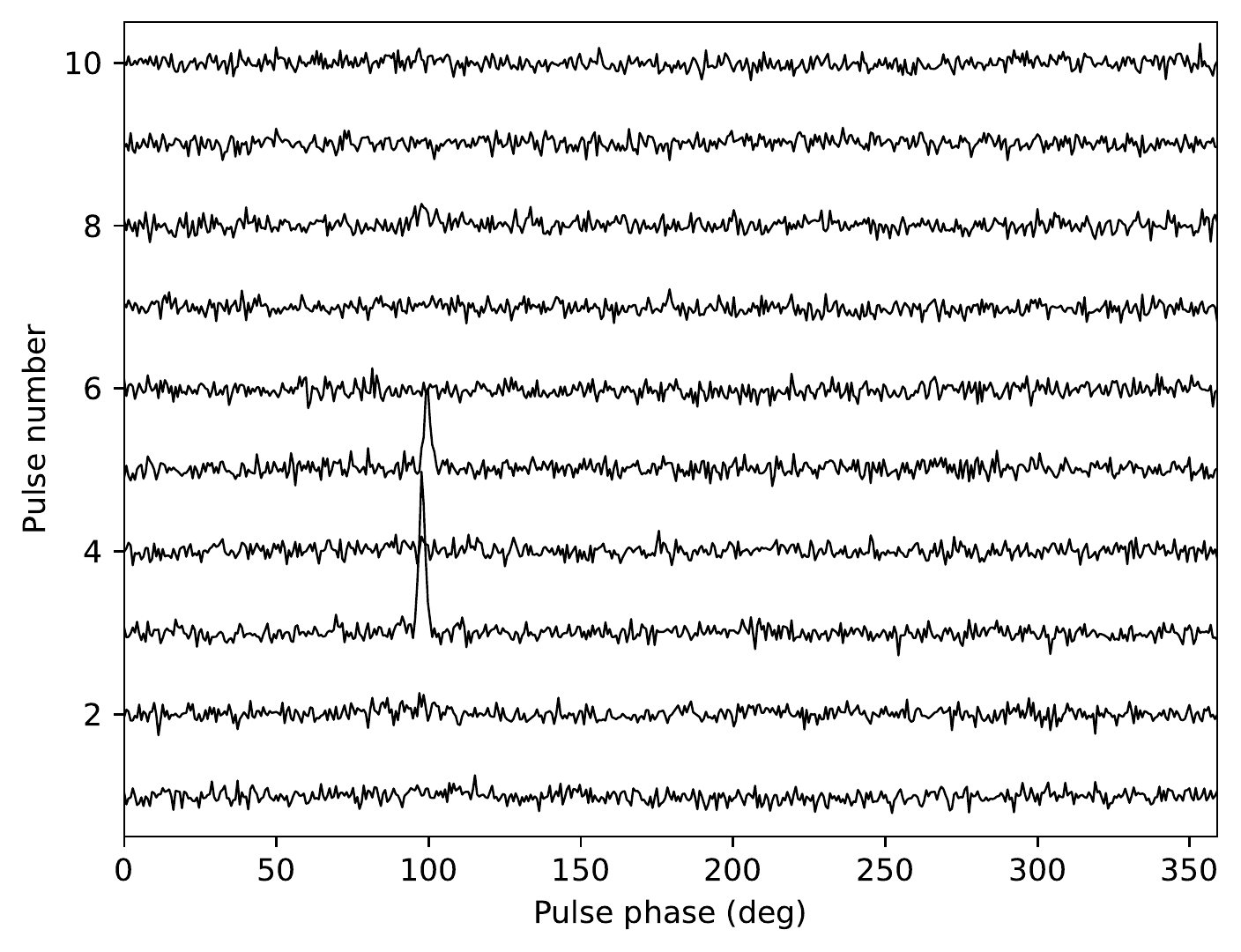}
\caption{A single-pulse stack of 10 single pulses for PSR~B1534+12.}\label{single}
\end{figure}

The 26115 individual pulses in our observation of PSR~B1534+12 is shown in the upper panel of Figure~\ref{gray} as a grey-scale image. We also display a pulse stack containing 10 adjacent individual pulses is shown in Figure~\ref{single}. 

The pulse profile can be divided into three components, two in the main pulse and the interpulse. A ``weak emission'' region is labeled as ``I'' in the bottom panel of Figure~\ref{gray}. A narrower, brighter component is labeled as ``II'' (note that region ``II'' does not include region ``I''). The interpulse is labeled as ``III''.

As seen in Figures~\ref{gray} and ~\ref{single}, the intensity in region ``II'' is highly variable and we term this as ``burst emission''.  The weak state is always present.  When both exist then the two states overlap. To identify whether the burst state exists in any particular pulse, we determine the signal to noise ratio (S/N) as $I_{\rm peak}/\sigma_{\rm off}$, where $I_{\rm peak}$ is the peak intensity of a single pulse in the region ``II'' and $\sigma_{\rm off}$ is the root mean square (rms) of the off pulse region. 
If the S/N for a single pulse is larger than $5\sigma$ then we initially classify it as likely being in the burst state.  We subsequently check every single pulse in this selection of pulses to confirm our classification.  In total 443 single pulses in the burst state were obtained using this method.  For pulses with the S/N smaller than $3\sigma$, we classify the pulse as belonging to the weak state. Between $3$ and $5\sigma$, we do not attempt to classify the state.

\begin{figure}
  \centering
  \includegraphics[width=80mm]{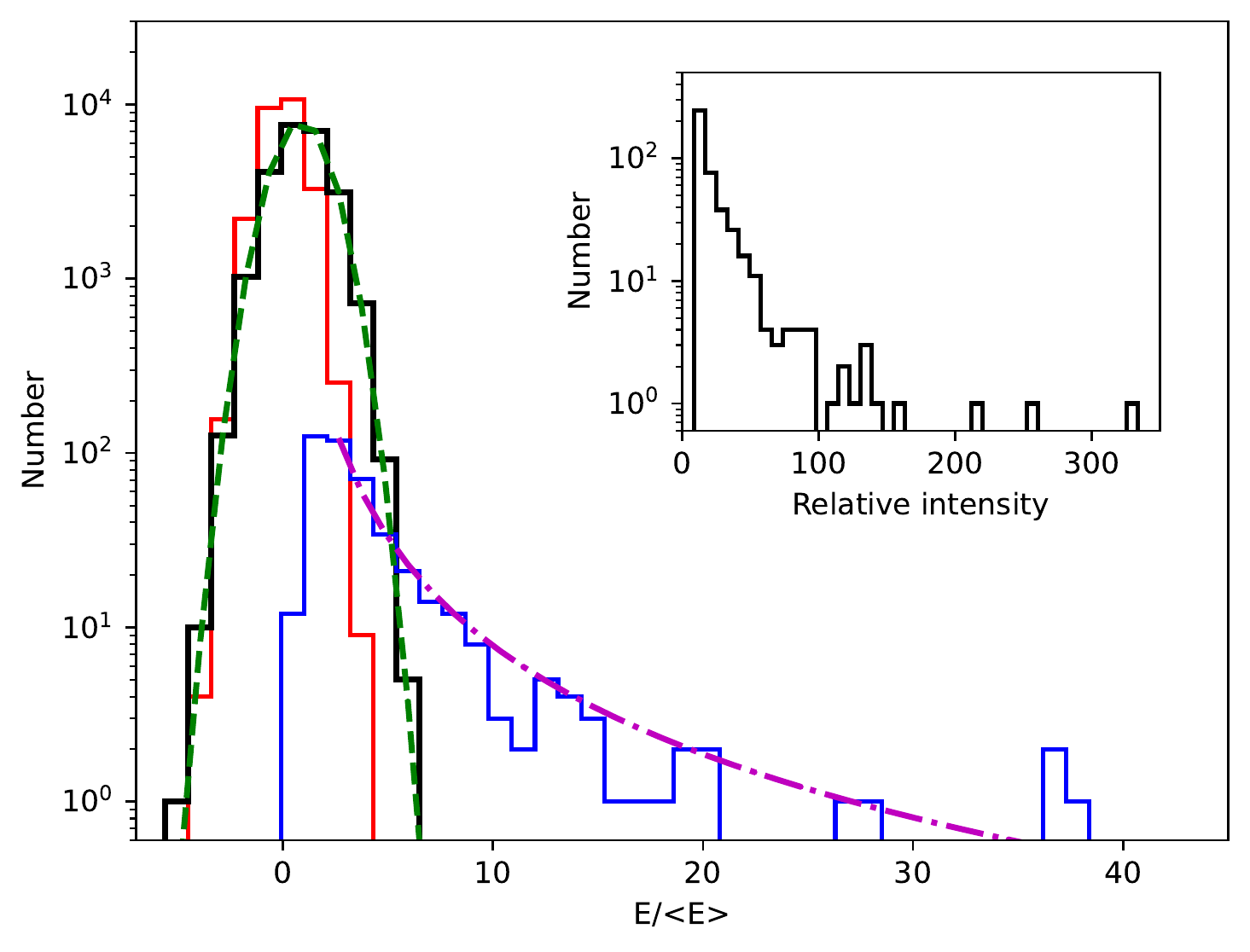}
  \caption{Pulse energy histograms for the off-pulse region (red solid line), weak state (black solid line) and burst state (blue solid line) for PSR~B1534+12. The pulse energies are normalised by the mean on-pulse energy. The green dash-dashed line is noise and weak state distribution.
  The magenta dash-dotted line is power-law distribution fitted to the burst states. The distribution of the relative peak flux intensity compared to the average pulse for the pulses in the burst state is shown in the inner plot.}\label{energy}
\end{figure}

To determine the pulse energy distributions we calculate the area (in non-physical units) across the regions ``I'' and ``II'' for the weak pulse and the regions ``II'' for the burst pulse, respectively.  We also calculate the area in an off-pulse region.   We normalise each pulse by the mean value in the on-pulse region.  The results are shown in Figure~\ref{energy}.  The noise in the off-pulse region (the red solid line) follows the expected normal distribution. For many of the pulses in the weak state, we cannot distinguish them from the off-pulse noise, but there is a significant excess at high energies and we fit this both with normal+normal  and normal+log-normal distributions. The two fits are almost identical and therefore we cannot distinguish between a normal or a log-normal distribution for the weak state pulses. The blue histogram shows the distribution of the burst emission and we model that using a power-law distribution with a power-law exponent of $-2.06\pm0.05$.  

The emission from this pulsar at 430\,MHz has a similar energy distribution (see Figure 1 in \citealt{Sallmen95}). However, the single pulse energy in our observation extended out to about 38 times the average pulse energy, which is much large than reported in the earlier paper at 430\,MHz (which extended to $\sim$13 times the average pulse energy).

\begin{figure}
  \centering
  \includegraphics[width=80mm]{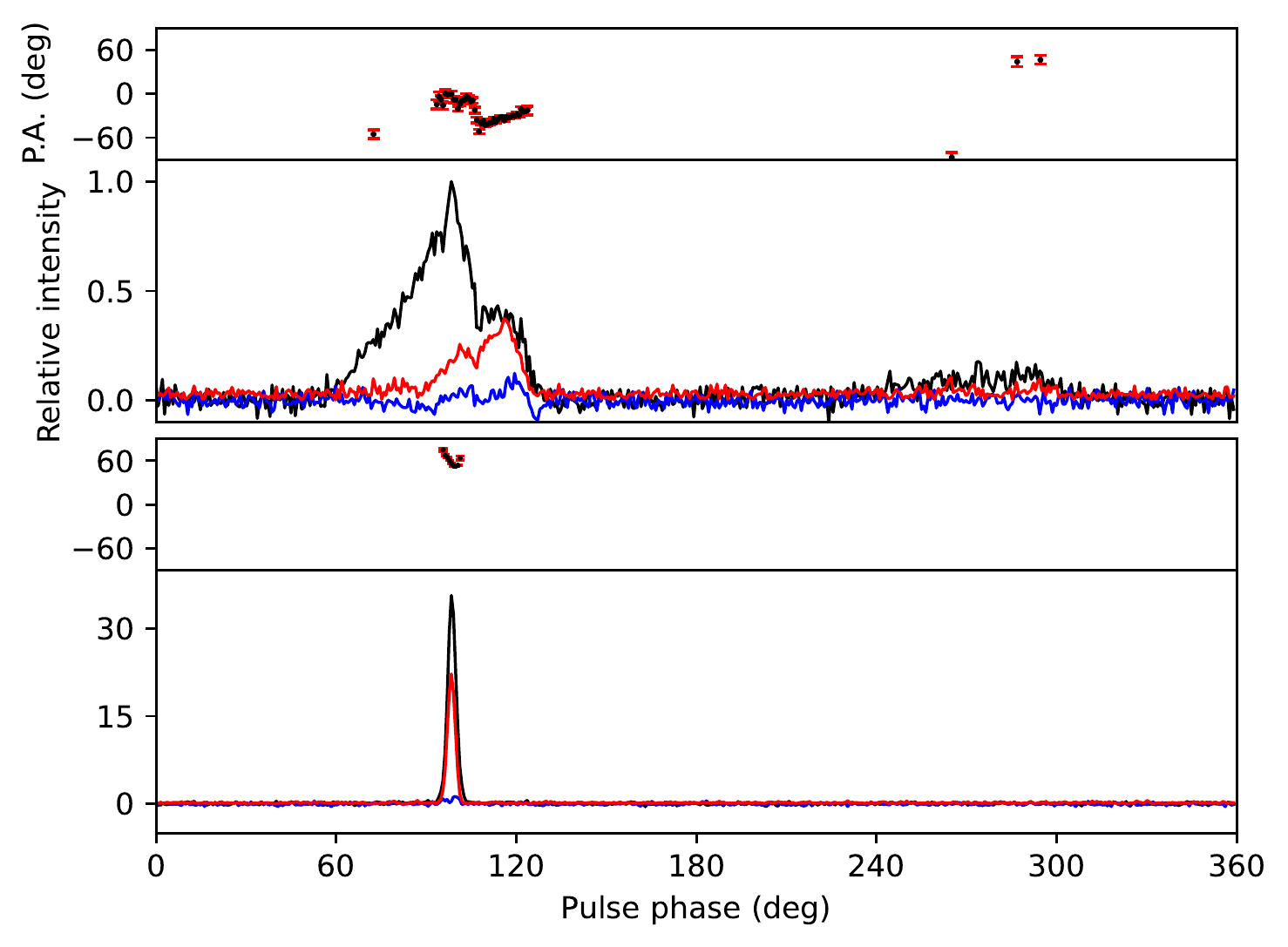}
  \caption{Polarisation profiles for PSR~B1534$+$12 in the weak (upper panel) and burst states (bottom panel). The position angle (black dots) and corresponding uncertainties (red bars) of the linearly polarised emission are shown as a function of pulse phase. The black, red and blue lines are for the total intensity, linearly polarised intensity and circularly polarised intensity, respectively. All pulse profiles are normalised by the average profile for the weak state. }\label{prof}
\end{figure}

The average pulse profiles of the two emission states are shown in the upper and bottom panels of Figure~\ref{prof}. We do not observe any increase in the strength of the interpulse during a burst state. The pulse emission for weak state is always present. The pulse profile for the burst emission can be obtained by subtracting the average profile of the weak state  from the burst profile.  This is shown in the bottom panel of Figure~\ref{prof}. This burst emission is much narrower than that for the weak state. The full widths at half maximum (W50) of the profiles are 0.35\,ms and 3.78\,ms, respectively.  To form the polarisation profiles we used the cataloged rotation measure for this pulsar of $10.6\pm2$\,rad\,m$^{-2}$ \citep{Weisberg04}\footnote{We also used our own data to determine the rotation measure and obtained $10\pm13$\,rad\,m$^{-2}$. This is not unexpected because we are only using the 16 minutes data which provides a S/N of the profile of 137.9. This  value is consistent, but has a much larger uncertainty than the cataloged value.}.

\cite{Arzoumanian96} showed, by fitting the rotating-vector model, that this pulsar is nearly an orthogonal rotator.  However,
the polarisation properties of the two emission states differ.  The linear and circular polarisation are shown in Figure~\ref{prof} as red and blue lines respectively. The position angle of the linear polarisation is shown in the upper sub-panels for each plot.
The position angle difference between the two states ranges from 56 to 91 degrees across the burst-state profile.
In the weak state, the linearly polarised intensity increases gradually over the pulse phase.  It is low in the leading pulse component, whereas it becomes much higher (about 85\%) in the trailing component of the main pulse. In the burst state, the linearly polarised intensity is about 62\% of total intensity.

For single pulses in the burst state, the peak flux intensities are in the range 9 to 334 times stronger than the peak flux intensity averaged over all pulses (see the inner plot in Figure \ref{energy}). These bright pulses are generated in a narrow phase range, and have W50 pulse widths in the range of 49 to 674\,$\mu$s (the characteristic width of these bright pulses at 430\,MHz is $\sim 160\,\mu$s; \citealt{Sallmen95}).  We also analysed the polarisation profile of the brightest single pulse. The fractional linear polarisation is high (83\% at the peak) and significant circular polarisation is detected (18\% at its maximum value). 

\subsection{Timing the brightest pulses}

Our results show that the brighter pulses are generated in a narrow pulse phase range. This suggests that that it may be possible to improve the timing precision (at least over short timescales) could be  if only the brightest pulses are selected.  To study this, we divided the observation into 16 segments each of $\sim$1\, minute duration and, within each segment, summed separately all the pulses as well as those only in the burst state (S/N $> 5$) and those only in the weak state (S/N $< 3$). Three analytic pulse templates were formed from these summed profiles.

\begin{figure}
  \centering
  \includegraphics[width=80mm]{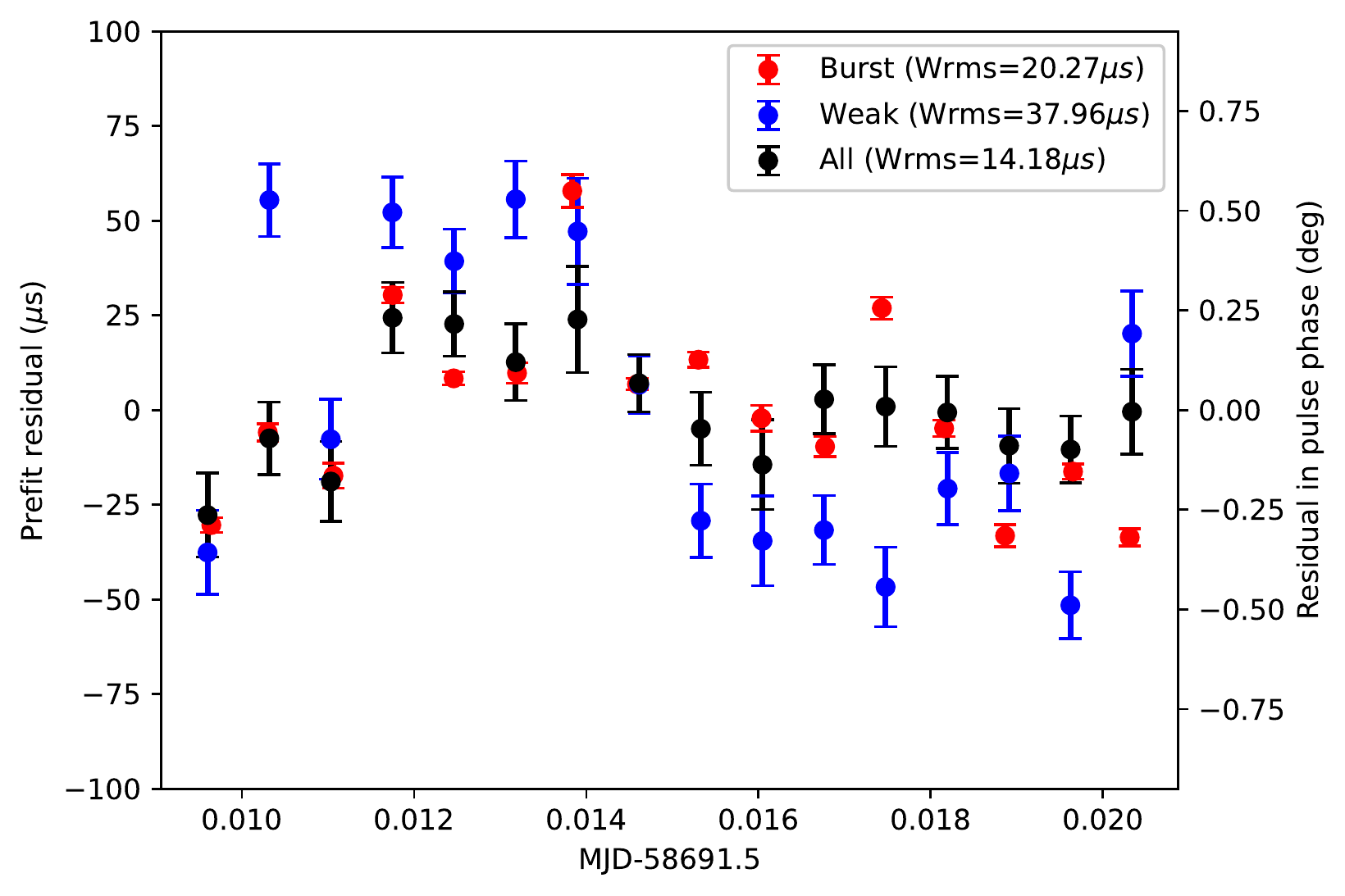}
  \caption{Timing residuals for the average pulse profiles formed by the pulses in burst state (red dots), weak state (blue dots) and all the pulses together (black dots). The bars give the ToA uncertainties. The weighted rms residuals are listed in the Figure legend. The mean ToA uncertainties are 2.5, 32.2 and 10.1\,$\mu s$, respectively.}\label{res}
\end{figure}

\begin{figure}
  \centering
  \includegraphics[width=80mm]{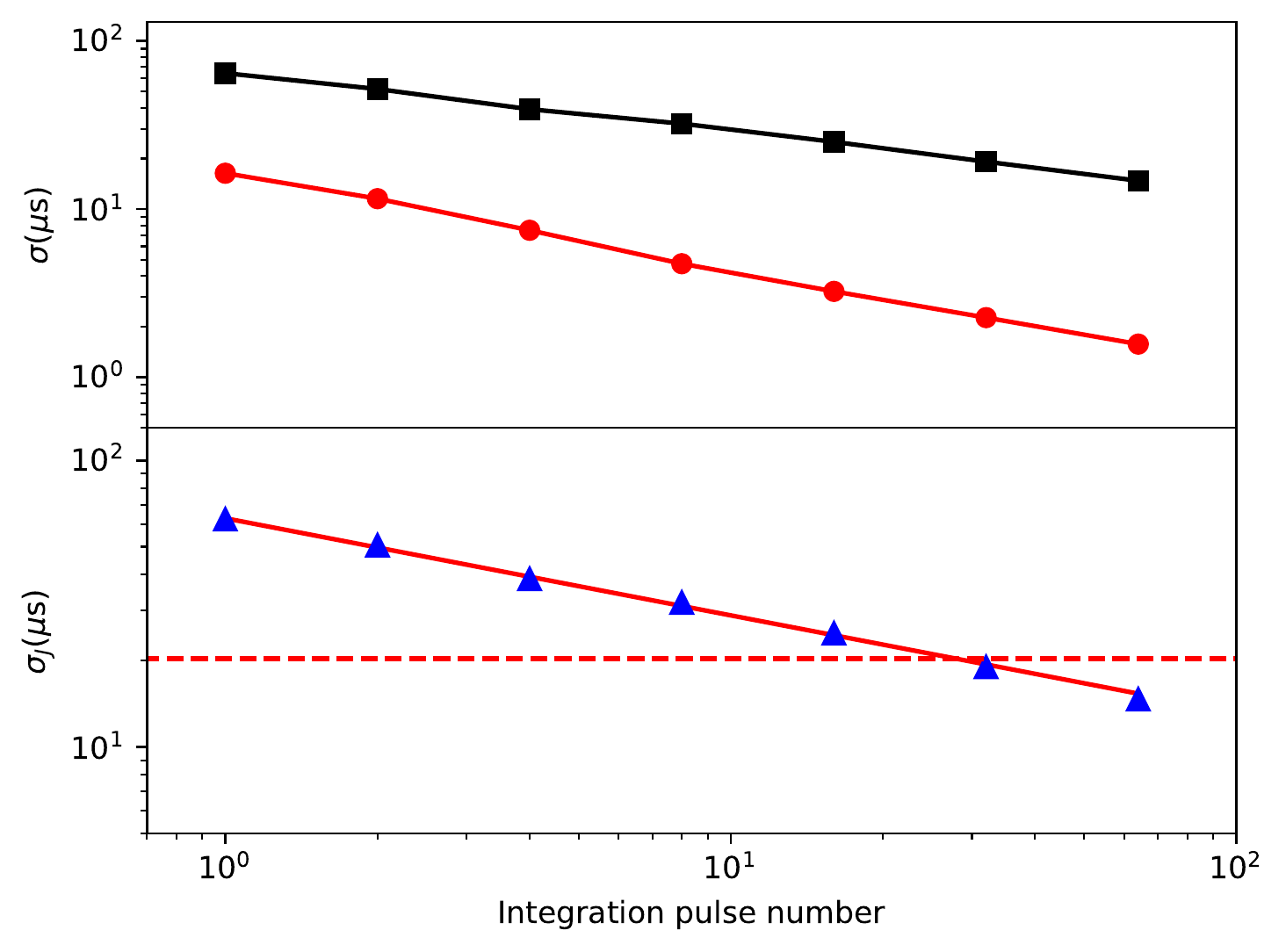}
  \caption{Determination of the jitter noise in PSR~B1534+12. Upper panel: the rms residuals versus the number of pulses averaged ($N_{\rm p}$)  (squares) and the average ToA uncertainties (circles). Bottom panel: the variations for $\sigma_{\rm J}{(N_{\rm p})}$ versus $N_{\rm p}$. The red line is the best fitting model for the jitter noise with the index of $-0.35\pm0.01$. The dashed line is for the weighted rms residual of 20.27\,$\mu s$.}\label{jitter}
\end{figure}

We formed pulse ToA by cross-correlating these resulting summed profiles using the relevant standard template. We then formed timing residuals using the timing model provided by the online pulsar catalogue~\citep{Manchester05}.   The timing residuals for the burst state, weak state and for all the single pulses are shown in Figure~\ref{res}. As expected, the ToA uncertainties were smaller for the pulses in the burst state. However, the timing residuals for the burst state scatter by significantly more than their uncertainties.   We quantify this jitter as the quadrature difference between the observed rms residual and the 
rms ToA uncertainty, $\sigma_{\rm J}^2{(N_{\rm p})}=\sigma_{\rm obs}^2{(N_{\rm p})}-\sigma_{\rm ToA}^2{(N_{\rm p})}$ where ${N_{\rm p}}$ is the number of pulses that have been averaged together.  Our results are shown in Figure~\ref{jitter} and the rms noise caused by this jitter per pulse, $\sigma_{\rm J}{(1)}= 62$\,$\mu s$.

\section{DISCUSSIONS AND CONCLUSIONS}

We have presented single pulse profiles for PSR B1534+12. The emission from this pulsar shows two different states, a weak state with a wide pulse profile and a burst state with a narrow pulse profile.
This phenomenon has never been seen in double neutron star systems before, but emission state changes like this have been observed in young pulsars~\citep{Weltevrede06}.
We note that this phenomenon is not like the typical mode changing which the pulsar emission switches between two or more stable states~\citep{Bartel82}. 
For PSR B1534+12, the weak emission is always present in our observation, while the burst emission appears in conjunction with the weak state, but only in a narrow pulse phase. These results suggest that the emission for the two states is generated in two different regions in the pulsar magnetosphere. This is evidenced by their different observation properties both in pulse energy distributions and polarization properties. Further multi-frequency observations (with instruments such as the Parkes ultra-wide-bandwidth receiver, see \citealt{Hobbs20}) are needed to further constrain the emission geometry of these regions. 

During the burst state we identified bright pulses which are similar to those defined as giant pulses. 
The bright state pulses follow a power-law distribution, which is the same as the pulse energy distribution for giant pulses~\citep{Johnston03,Karuppusamy10}.  However, this power-law distribution is detectable to relatively low pulse energies.  Pulsars that exhibit giant pulses are divided into two classes based on the magnetic field in the light cylinder $B_{\rm LC}$. For the first class, this is $\sim 10^6\,{\rm G}$ and that of the second class is from 10 to 100$\,{\rm G}$~\citep{Kuzmin04}.
The typical pulse width for the first class is measured in micro-seconds or even nano-seconds. The typical widths for the second class are several milliseconds.
The pulse widths of the bright pulses of PSR~B1534+12 lie in the range of several tens to several hundred micro-seconds, which is between the typical pulse width of the two giant pulses classes.
The $B_{\rm LC}$ of PSR~B1534+12 is about $1.67\times10^{3}\,{\rm G}$ and also falls into the gap between these classes.
This pulsar provides evidence for connection between the two giant pulse classes.

The bright pulses in the pulsar share some similar properties to the giant pulses of PSR~B1937+21. For instance, they are all generated in a narrow phase range. However, the typical pulse width for the bright pulses in PSR B1534+12 is much wider (the giant pulses for PSR~B1937+21 are typically a few hundred nanoseconds wide; \citealt{McKee19}). The bright pulses in PSR~B1534+12 exhibit strong linear polarization, whereas the giant pulses of PSR~B1937+21 are almost $100\%$ circularly polarized~\citep{Soglasnov04}.

The pulses in the burst state for PSR B1534+12 share some properties with the giant micro-pulses.  For instance, they follow power-law distributions and have the similar pulse widths.
However, they are different. Firstly, the giant micro-pulse is typically seen in young pulsars~\citep{Johnston02}, while PSR B1534+12 is a partly recycled pulsar. Secondly, the giant micro-pulse is a separate component of a single pulse, which may be caused by micro-beams and some are quasi-periodic~\citep{Kramer02}.  For PSR~B1534+12, our results suggest that the bright pulse emission is generated in a different region in the pulsar magnetosphere compared with the weak emission, which is always present.

Pulsars like PSR~B1534+12 can be used to study fundamental physics such as theories of gravity. Therefore, they deserve detailed studies of their emission properties and how those properties may affect the long-term timing campaigns. We studied the timing precision achievable for PSR~B1534+12 using the bright pulses and found that even though the ToA uncertainties decrease significantly when only these bright and narrow pulses are selected, the rms timing residuals (and hence the precision achievable for any relativistic study) for PSR~B1534+12 does not improve.  This is not unexpected as many pulsars exhibit jitter, e.g., PSR J0437$-$4715~\citep{liu12}, PSR J1713+0747~\citep{Shannon12} and PSR J1022+1001~\citep{Liu15}, and this is a major limitation of high precision pulsar timing experiments with sensitive telescopes (see, e.g., \citealt{Hobbs19}). 

We will continue to monitor this pulsar with FAST and expect that our results will provide more information on the single pulse statistics of this pulsar as well as the long-term timing behaviour.

\section*{Acknowledgments}
This work made use of the data from FAST (Five-hundred-meter Aperture Spherical radio Telescope).  FAST is a Chinese national mega-science facility, operated by National Astronomical Observatories, Chinese Academy of Sciences. This work is supported by the Youth Innovation Promotion Association of Chinese Academy of Sciences,  201$^*$ Project of Xinjiang Uygur Autonomous Region of China for Flexibly Fetching in Upscale  Talents, the National Key Research and Development Program of China (2016YFA0400804), the Operation, Maintenance and Upgrading Fund for Astronomical Telescopes and Facility Instruments, budgeted from the Ministry of Finance of China (MOF) and administrated by the Chinese Academy of Science (CAS) and the Key Lab of FAST, National Astronomical Observatories, Chinese Academy of Sciences.

\end{document}